\documentclass[
 aps,prb,
 amsmath,amssymb,
reprint]{revtex4-1}

\usepackage{amssymb}
\usepackage{graphicx}
\usepackage{amsmath}
\usepackage{longtable}

\begin{document}

\title{A One-Step Model of Photoemission from Single Crystal Surfaces}

\author{Siddharth Karkare}
\affiliation{Lawrence Berkeley National Laboratory, 1 Cyclotron Rd., Berkeley, CA, USA
94720}

\author{Weishi Wan}
\affiliation{Lawrence Berkeley National Laboratory, 1 Cyclotron Rd., Berkeley, CA, USA
94720}

\author{Jun Feng}
\affiliation{Lawrence Berkeley National Laboratory, 1 Cyclotron Rd., Berkeley, CA, USA
94720}

\author{Tai C. Chiang}
\affiliation{Department of Physics, University of Illinois, Urbana, IL 61801 USA}
\affiliation{Frederick Seitz Materials Research Laboratory, University of Illinois,
Urbana, IL 61801 USA}

\author{Howard A. Padmore}
\affiliation{Lawrence Berkeley National Laboratory, 1 Cyclotron Rd., Berkeley, CA, USA}

\begin{abstract}
In this paper, we present a 3-D one step photoemission model that can be used to calculate the quantum efficiency and momentum distributions of electrons photoemitted from ordered single crystal surfaces close to the photoemission threshold. Using Ag(111) as an example, we show that the model can not only calculate the quantum efficiency from the surface state accurately without using any ad hoc parameters, but also provides a theoretical quantitative explanation of the vectorial photoelectric effect. This model in conjunction with other band structure and wave function calculation techniques can be effectively used to screen single crystal photoemitters for use as electron sources for particle accelerator and ultrafast electron diffraction applications.  \end{abstract}

\maketitle

\section{Introduction}

Over the past few decades photoemission based tools like Photoelectron Spectroscopy (PES) and Angle-Resolved Photoelectron Spectroscopy (ARPES) have proven extremely successful in studying the chemical and electronic structure of solid state materials and surfaces\cite{Kevan1992}. As a result, physics of the photoemission phenomena has been well investigated with regards to explaining the angle resolved electron energy spectra obtained using UV and X-ray light sources.

More recently, photoemission has gained popularity as a source of electrons for several applications like Free Electron Lasers (FELs)\cite{FEL} and Ultrafast Electron Diffraction (UED)\cite{UED} experiments. The quantum efficiency (QE) and the transverse (to the normal on the photoemission surface) momentum spread or the rms transverse momentum are the most critical figures of merit of the photoemission based electron sources (or photocathodes) that limit the performance of such applications\cite{NIMA}. For example, the transverse coherence length of the electron beam in UED which limits the largest lattice size that can be studied is inversely proportional to the rms transverse momentum of electrons emitted from the cathode\cite{karkare_prl}. The transverse momentum spread also limits the smallest possible electron beam emittance which defines the shortest possible lasing wavelength of an FEL\cite{FEL_book}. The QE determines the drive laser power needed to obtain the electron bunch charge required for the particular application; a low QE can implies high drive laser power often making the drive laser system prohibitively complex and expensive\cite{LCLS_laser}. High drive laser power can also limit the smallest possible rms transverse momentum through ultrafast laser heating of the electron gas\cite{ultrafast_heating}. Hence a high QE is required.     

Despite the technological importance of solid state photoemission as an electron source, the physics that governs the relevant photoemission properties of QE and rms transverse momentum is not well understood. The first theory to model these photoemission parameters from metal cathodes was proposed by Dowell and Schmerge and followed a three step photoemission model\cite{Dowell}. This theory successfully explained the QE and rms transverse momentum obtained from polycrystalline or disordered cathodes but did not model photoemission very close to the threshold accurately. An extension to this theory was developed recently, to model photoemission near the threshold\cite{Jun_thermal}. It showed that the smallest possible rms transverse momentum from polycrystalline surfaces is thermally limited by the temperature of the lattice. However these models did not include the effects of band structure, polarization and angle of incident light (the vectorial photoelectric effect\cite{Broudy1971,Pedersoli2008} and did not model emission from single crystal surfaces of metals. A technique to estimate the rms transverse momentum spread from single crystal faces of metal cathodes using the band structure calculated from density functional theory was developed by Schroeder \textit{et al}\cite{Andreas}. However, this technique does not estimate the QE and assumes uniform probability of photoemission from any given electron state which is generally not true.

In this paper, we present a scheme to calculate the QE and transverse momentum spread accurately using an one-step photoemission model. Our model is a 3-D expansion of the 1-D photoemission model developed by Miller \textit{et al.}\cite{Miller1996,Miller1997} to explain ultraviolet photoemission spectra (UPS) obtained from single crystals surfaces of noble metals. Photoemission is modelled as a one step process of the transition of electrons from the initial bulk or surface state (ss) inside the metal to a time-reversed LEED like free electron state under the influence of the electromagnetic field of the incident light. We obtain the rate of such a transition using the Fermi golden rule to calculate the QE and the rms transverse momentum of emitted electrons. This photoemission model takes into account the effects of band structure, polarization and angle of incident light. 

Using the (111) surface of silver as an example, we show that our model predicts the QE near threshold accurately and explains the effects of polarization of incident light and angle of incidence quantitatively. We show that the emission from an Ag(111) surface at threshold, at an angle of incidence near 60$^{\circ}$ is dominated by the electrons emitted from the Shockley surface state\cite{Shockley1939} resulting in a QE of greater than $5\times10^{-5}$ very close to the photoemission threshold.  We also calculate the rms transverse momentum of electrons emitted from the Ag(111) surface. We show that the Ag(111) surface can simultaneously provide a high QE and a low rms transverse momentum very close to the thermal limit\cite{Jun_thermal} and hence can be used as an excellent electron source.

The dependence of QE on the polarization of incident light and angle of incidence is called the vectorial photoelectric effect and has been investigated experimentally, but has been modeled only empirically\cite{Broudy1971,Pedersoli2008}. Using our scheme to calculate the QE we show that the vectorial photoelectric effect results from the variation of the overlap integral with the angle of incidence and polarization of incident light and can be modeled without use of any empirical data.

\section{The One-Step Model}
\subsection{Basic Formalism}

We assume that the normal to the solid-vacuum interface is along the $z$ direction and the classical interface is located at $z=0$, with $z>0$  being the vacuum side. 
The Hamiltonian of the photon-electron interaction is given by
\begin{eqnarray}
\mathcal{H} &=&\frac{1}{2m_{e}}\left( \vec{p}-\frac{e}{c}\vec{A}\right) ^{2}-
\frac{p^{2}}{2m_{e}} \\
&\mbox{\hspace{0.3mm}}\approx&-\frac{e\hbar }{m_{e}c}\vec{A}\cdot \vec{\nabla}-
\frac{e\hbar }{2m_{e}c}\left( \vec{\nabla}\cdot \vec{A}\right) \label{eq:hamiltonian}
\end{eqnarray}
where $\vec{p}$ is the momentum operator, $\vec{A}$ is the vector potential of the incident light, $e$ is the unit charge, $c$ is the speed of light and  $m_e$ is the mass of a free electron in vacuum.

The vector potential of incident light inside the metal surface can be given by $\vec{A}=A_0e^{\frac{z-z_0}{d_l}}\vec{\epsilon}$, where $A_0$ is the magnitude of the incident vector potential just outside the surface, $\vec{\epsilon}$ is the polarization vector inside the surface, $d_l$ is the decay length of the incident light in the metal and $z_0$ is the location of the interface adjusted to account for the spilling over of the electron cloud into vacuum\cite{Miller1996,Miller1997} due to the surface state. For the Ag(111) surface $z_0$ is determined by wave function matching of the Shockley surface state at the solid-vacuum interface as shown in section III B. Note that the magnitude of polarization vector $\vec{\epsilon}$ is not unity and takes into account the reflection at the surface as given in section II B. The incident photon flux per unit area is given by
\begin{equation}
F=\frac{2\varepsilon_0\left\vert A_{0}\right\vert ^{2}\omega}{\hbar c }\cos\left(\theta_i\right)\label{flux}
\end{equation}where $\omega$ is the frequency of incident light, $\varepsilon_0$ is the dielectric constant of vacuum and $\theta_i$ is the angle of incidence\cite{sakurai}. 

For ultra-violet light, the wavelength is long enough that the $\vec{\nabla}\cdot\vec{A}$ term in equation \ref{eq:hamiltonian} can be ignored everywhere except at the metal-vacuum interface. At the metal-vacuum interface, there is a sharp discontinuity in $\vec{A}$ in the $z$ direction and $\vec{\nabla}\cdot\vec{A}$ results in a delta function at $z=z_0$. The hamiltonian is then given by
\begin{equation}  
\mathcal{H}=-\frac{e\hbar A_0e^{\frac{z^{\prime }H\left(-z^{\prime }\right)}{d_l}} }{m_{e}c}\left(\vec{\epsilon}\cdot \vec{\nabla}+C\epsilon_z\delta\left(z^{\prime }\right)\right)\label{hamiltonian}
\end{equation}
where $\epsilon_z$ is the $z$ component of $\vec{\epsilon}$, $H\left(z\right)$ is the Heaviside function, $z^{\prime }=z-z_0$ and C is a constant that depends only on the photon energy and the properties of the solid. The constant C can be obtained by fitting the calculations of the 1D model to the photoemission electron spectroscopy data\cite{Miller1996,Miller1997}.

Photoemission from single crystal surfaces can be modeled as a transition process of an electron between an initial bulk or surface state (ss) inside the lattice with wave function $\phi_i$ to a time reversed LEED like free electron state in vacuum with wave function $\phi_f$ under the influence of incident light\cite{Mahan1970,Kevan1992}.
The total transition rate of this process is given by Fermi's golden rule as 
\begin{equation}
R=\sum_{i}\sum_{f}\frac{4\pi }{\hbar }\left\vert \left\langle \phi _{f}\left\vert \mathcal{H}\right\vert \phi _{i}\right\rangle \right\vert ^{2}\delta\left(E_f-\left(E_i+\hbar\omega\right)\right)F\left(E_i\right) ,
\label{trate}\end{equation}
where the summations are over all possible initial and final states, $E_i$ and $E_f$ are the energies of the initial and final states respectively, the $\delta$ function enforces the conservation of energy and  $F\left(E_i\right)=\left(1+\exp\left({\frac{E_{i}}{k_{B}T}}\right)\right)^{-1}$ is the Fermi-Dirac distribution. $k_B$ is the Boltzmann constant and $T$ is the temperature of the lattice. Note that we have assumed the Fermi level to be 0. The expression for the transition rate includes a factor of 2 to account for the two possible electron spins.

 We work within the box approximation to assume that the volume under consideration extends from $-L/2$ to $L/2$ in all directions and L$\rightarrow\infty$. Within this assumption we can convert the summations in equation \ref{trate} to integrals and rewrite the transition rate as
\begin{equation}
R=\frac{4\pi}{\hbar}\left(\frac{L}{2\pi}\right)^6\int d^3\vec{k_i}\int d^3\vec{k}M ^{2}\delta\left(E_f-\left(E_i+\hbar\omega\right)\right)F\left(E_i\right)\label{trate1}
\end{equation}
where $M=\left\vert \left\langle \phi _{f}\left\vert \mathcal{H}\right\vert \phi _{i}\right\rangle \right\vert$ is the overlap integral or the matrix element. $\vec{k_i}=k_{ix}\hat{x}+k_{iy}\hat{y}+k_{iz}\hat{z}$ is the wave vector of electrons in their initial state and $\vec{k}=k_{x}\hat{x}+k_{y}\hat{y}+k_{z}\hat{z}$ is the wave vector of the emitted electron. If the work function of the emission surface is $W$, the energy of the final state, $E_f$, can be written as 
\begin{equation}
E_f=\frac{\hbar^2k_{z}^2}{2m_e}+\frac{\hbar^2k_{r}^2}{2m_e}+W,
\end{equation} 
where $k_{r}=\sqrt{k_x^2+k_y^2}$ is the wave number of the emitted electron in the transverse direction ($x$-$y$ plane). The delta function in equation \ref{trate1} can then be written as
\begin{equation}
\delta\left(E_f-\left(E_i+\hbar\omega\right)\right)=\frac{\sqrt{m_e}}{\hbar\sqrt{2X}}\delta\left(k_z-k_{z0}\right)\label{delta}
\end{equation}
where $k_{z0}=\frac{\sqrt{2m_eX}}{\hbar}$, $X=E_i+\hbar\omega-\frac{\hbar^2k_{r}^2}{2m_e}-W$. 

The QE can simply be calculated as
\begin{equation}
\mathrm{QE}=\frac{R}{FL^2}
\end{equation}
As will be shown in the following section, the QE is independent of $L$ as the matrix element is proportional to $1/L^2$ owing to the normalization of the wave functions within the bounding box.

The transverse momentum spread or the rms transverse momentum can be calculated as
\begin{equation}
 \sqrt{\langle p_{r}^2\rangle}=\left[\frac{\int d^3k_i\int d^3k_f\hbar^2k_r^2M ^{2}\delta\left(E_f-\left(E_i+\hbar\omega\right)\right)F\left(E_i\right)}{\int d^3k_i\int d^3k_fM ^{2}\delta\left(E_f-\left(E_i+\hbar\omega\right)\right)F\left(E_i\right)}\right]^{\frac{1}{2}}\label{rmsmom}
 \end{equation}
 
One can obtain the QE and the rms transverse momentum by calculating the matrix elements $M$ and evaluating the integrals in equation \ref{trate1}. Calculation of the matrix elements cannot be generalized further and requires the knowledge of the band structure, wave functions and the orientation of the photo-emitting surface. In the next section, we calculate the matrix elements and perform the integrals to obtain the QE and the rms transverse momentum for the Ag(111) surface as an example.  

\subsection{Refraction of Light at the Solid-Vacuum Interface}
In order to calculate the matrix elements, one needs to obtain the polarization vector ($\vec{\epsilon}$) for incident light inside the solid surface.
Expressions to obtain $\vec{\epsilon}$ can be found in Born and Wolf's Principles of Optics\cite{bornwolf} or any other standard text on electromagnetic waves. However, we state them here for the sake of completion.

We assume $x$-$z$ plane to be the plane of incidence.
The complex angle of transmission is given by Snell's law as
\begin{equation}\theta_{t}=\arcsin{\left(\frac{1}{n}\sin{\left(\theta_i\right)}\right)}\end{equation}
where $n=n_r+in_i$ is the complex index of refraction, $\theta_i$ is the angle of incidence,

The angle of the light wave vector inside the metal with respect to the $z$ axis can be given by
\begin{equation}
\theta _{t}^{^{\prime }}=\arctan \left[ \frac{\sin \theta _{i}}{q\left(
n_r\cos \gamma -n_i\sin \gamma \right) }\right] 
\end{equation}
and the optical decay length for the fields can be given by
\begin{equation}
d_{l}=\frac{c}{\omega q\left( n_i\cos \gamma +n_r\sin \gamma \right) }. 
\end{equation}
where
\begin{equation}
q=\left[\left[ 1-\frac{n_r^{2}-n_i^{2}}{\left( n_r^{2}+n_i^{2}\right) ^{2}}\sin
^{2}\theta _{i}\right] ^{2}+\left[ \frac{2n_rn_i}{\left( n_r^{2}+n_i^{2}\right) ^{2}}%
\sin ^{2}\theta _{i}\right] ^{2}\right]^{\frac{1}{4}} 
\end{equation}
and
\begin{equation}
\gamma =\frac{1}{2}\arctan \frac{2n_rn_i\sin ^{2}\theta _{i}}{\left(
n_r^{2}+n_i^{2}\right) ^{2}-\left( n_r^{2}-n_i^{2}\right) \sin ^{2}\theta _{i}}. 
\end{equation}

For $p$-polarized light the polarization vector of the vector potential is 
\begin{equation}
\vec{\epsilon}=T_{p}\sin \theta_{t}\hat{z}+T_{p}\cos \theta_{t}\hat{x},\end{equation}
where \begin{equation}T_p=\frac{2\cos \theta _{i}}{\cos \theta _{t}+n\cos \theta _{i}}.\end{equation}  

For $s$-polarized light the polariztion vector of the vector potential is 
\begin{equation}
\vec{\epsilon}=T_{s}\hat{y}, 
\end{equation}
where \begin{equation}T_s=\frac{2\cos \theta _{i}}{n\cos \theta _{t}+\cos \theta _{i}}.\end{equation}

\section{Photoemission from A\lowercase{g}(111)}

In this section, we demonstrate the use of the formalism developed above to obtain analytic expressions for the QE and rms transverse momentum from a Ag(111) surface. The calculated QE matches the experimental values showing the effectiveness of the formalism developed above.

\subsection{Band structure of Ag(111)}
We use  a two-band fit to the nearly free-electron like Ag $sp$ band dispersion model around the $L$ point\cite{Smith1985}.
The total energy ($E_{i,f}$) can be divided into the longitudinal part ($E_{zi,zf}$) and the transverse part ($E_{ri,rf}$) and can be written as
\begin{equation}
E_{i,f}=E_{zi,zf}+E_{ri,rf}
\end{equation}

\begin{figure}
\includegraphics[width=3.2in]{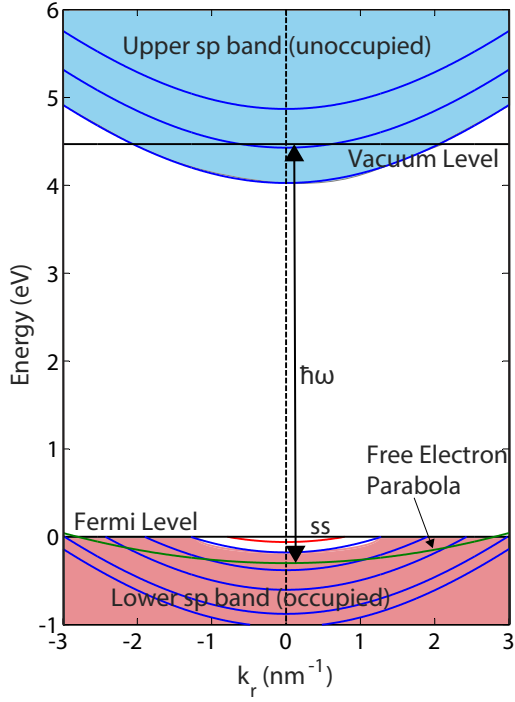}
\caption{Band structure of Ag (within the nearly free electron model) projected onto the [111] direction. The pink shaded region is the lower $sp$ bands filled with electrons. The region shaded in blue is the upper $sp$ bands which is unoccupied. The solid red line is the surface state (ss). The solid blue curves are contours of constant $E_{zi,zf}$ or correspondingly constant $k_{zi,zf}$. The green curve is the 'free electron parabola' corresponding to a photon energy of $\hbar\omega$. The conservation of energy and transverse momentum allow only the occupied states above this parabola to be emitted.
\label{fig1}}
\end{figure}

\subsubsection{Band structure of bulk states}  
Within the framework of the nearly free electron model, the dispersion relations for the two bands in the longitudinal direction ([111] or $z$ direction) are given by\cite{Chaing2000} 
\begin{equation}
E_{zi,zf}=E_{v}+V+\frac{\hbar^2k_{zi,zf}^2}{2m_{i,f}}\mp\left(\frac{\hbar^4p^2k_{zi,zf}^2}{m_{i,f}^2}+V^2\right)^{1/2}\label{dispersion}
\end{equation}
where $E_{zi,zf}$ are the longitudinal energies of the electrons in the lower and upper $sp$ bands respectively, $E_v$ is the valance band maximum, $V$ is the absolute value of the pseudo potential form factor and equals one-half of the gap at the zone boundary, $p$ is the magnitude of the wave vector at $L$ point and is equal to $\frac{\sqrt{3}\pi}{a}$ ($a$ being the lattice constant) and $m_{i,f}$ are the effective mass parameters of the lower and upper $sp$ bands respectively. It should be noted that $m_{i,f}$ represent higher order corrections from multi band effects and do no correspond to the curvature of the dispersion relations. The subscripts $i$ and $f$ represent the lower and upper $sp$ bands respectively. The Fermi level is assumed to be 0. The scale for $k_{zi,zf}$ is chosen such that the zero lies at the $L$ point.

The dispersion relation in the transverse directions ($x$-$y$ plane) is assumed to be cylindrically symmetric and can be modeled by nearly parabolic bands given by
\begin{equation}
E_{ri,rf}=\frac{\hbar^2k_{ri,rf}^2}{2m_{ri,rf}}+\sum_{n=3}^6\eta_{ni,nf}k_{ri,rf}^n\label{trans_bstruct}
\end{equation}  
where $E_{ri,rf}$ are the transverse energies of the electrons in the lower and upper $sp$ bands respectively, $k_{ri,rf}^2=k_{xi,xf}^2+k_{yi,yf}^2$ and $m_{ri,rf}$ are the transverse effective masses of the lower and upper $sp$ bands respectively. The coefficients $\eta_{ni,nf}$ are higher order correction coefficient obtained by fitting the band structure of silver\cite{Goldmann2003}.

Figure \ref{fig1} shows the band structure of Ag projected along the [111] direction. The pink shaded region is the lower $sp$ bands filled with electrons. The states of this band that extend beyond the Fermi level are unoccupied and not shown. The region shaded in blue is the upper $sp$ bands which is unoccupied. The solid blue curves are contours of constant $E_{zi,zf}$ or correspondingly constant $k_{zi,zf}$. The shape of these contours is nearly parabolic and given by equation \ref{trans_bstruct} with an offset of $E_{zi,zf}$.

Values of all parameters used for modeling the bulk band structure are given in table \ref{tab1} and were obtained by fitting the band structure of silver\cite{Goldmann2003}.

\subsubsection{Band structure of surface state}
Ag(111) exhibits a Shockley surface state\cite{Shockley1939} within the band gap at the $L$ point with energy $E_s$. The surface state has $E_{zi}=E_s$. Since the surface state is located within the band gap, $k_{zi}$ obtained from equation \ref{dispersion} is imaginary for the surface state\cite{hufner_book}. The dispersion relation in the transverse direction is parabolic and given by 
\begin{equation}
E_{ri}=\frac{\hbar^2k_{ri}^2}{2m_{s}}
\end{equation}

The effective mass of the surface state has been measured to be $m_s=0.40m_e$\cite{hufner1}. The energy of the surface state $E_s$ can change significantly with the sample and surface preparation methods and is sensitive to the strain in the crystal. At room temperature it has been reported to range between -20 meV to -120 meV\cite{hufner2,temp_dep}. Here, we use it as a fitting parameter and obtain the best fit for QE at $E_s=-100$ meV.

\begin{longtable*}{ | p{.15\textwidth} | p{.70\textwidth} | p{.15\textwidth} |}

     \hline
    $\mathrm{\mathbf{Symbol}}$ & $\mathrm{\mathbf{Description}}$ & $\mathrm{\mathbf{Value}}$ \\ \hline 
    $m_{i}$ & Longitudinal effective mass parameter of the lower $sp$ band & $0.80m_e$ \\ \hline
    $m_{f}$ & Longitudinal effective mass parameter of the upper $sp$ band & $0.90m_e$\\  \hline
$m_{ri}$ & Transverse effective mass of lower  $sp$ band & $0.35m_e$\\  \hline
$m_{rf}$ & Transverse effective mass of upper  $sp$ band & $2.60m_e$\\  \hline
$m_{s}$ & Effective mass of surface state & $0.40m_e$\\  \hline
$\eta_{3i}$ & Third order correction coefficient for lower $sp$ band&$-0.8\times10^{-3}$ eVnm$^{3}$\\  \hline
$\eta_{4i}$ & Fourth order correction coefficient for lower $sp$ band&$-1.0\times10^{-3}$ eVnm$^{4}$\\  \hline
$\eta_{5i}$ & Fifth order correction coefficient for lower $sp$ band&$-3.5\times10^{-5}$ eVnm$^{5}$\\  \hline
$\eta_{6i}$ & Sixth order correction coefficient for lower $sp$ band&$12.5\times10^{-6}$ eVnm$^{6}$\\  \hline
$\eta_{3f}$ & Third order correction coefficient for upper $sp$ band&$2.2\times10^{-3}$ eVnm$^{3}$\\  \hline
$\eta_{4f}$ & Fourth order correction coefficient for upper $sp$ band&$-6.5\times10^{-4}$ eVnm$^{4}$\\  \hline
$\eta_{5f}$ & Fifth order correction coefficient for upper $sp$ band&$-5.6\times10^{-5}$ eVnm$^{5}$\\  \hline
$\eta_{6f}$ & Sixth order correction coefficient for upper $sp$ band&$20.8\times10^{-6}$ eVnm$^{6}$\\  \hline
$E_s$ & Energy of surface state&$-100$ meV\\  \hline
$E_v$ & Valance band maximum at L point &$-178$ meV\\  \hline
$V$ & Pseudo potential form factor (equals one half band gap at L point) &$2.1$ eV\\  \hline
$W$ & Work function of Ag(111) &$4.45$ eV\\  \hline
$a$ & Unit cell length &$0.409$ nm\\  \hline
    \caption{\label{tab1}List of symbols and values used to model the band structure of Ag(111) surface}
\end{longtable*}

\subsection{Wave functions}

Close to the $L$ point, the $x$ and $y$ dependent part of the initial and final wave functions can be expressed as plane waves. Thus the initial and final wave functions can be expanded as $\phi_i=\phi_{zi}e^{ik_{xi}x}e^{ik_{yi}y}$ and $\phi_{f}=\phi_{zf}e^{ik_{xf}x}e^{ik_{yf}y}$ respectively. In order to match the transverse part of the final wave functions at the boundary we require $k_{xf}=k_x$ and $k_{yf}=k_y$. Below we give the $z$ dependent parts of the wave functions.

\begin{figure*}
\includegraphics[width=7in]{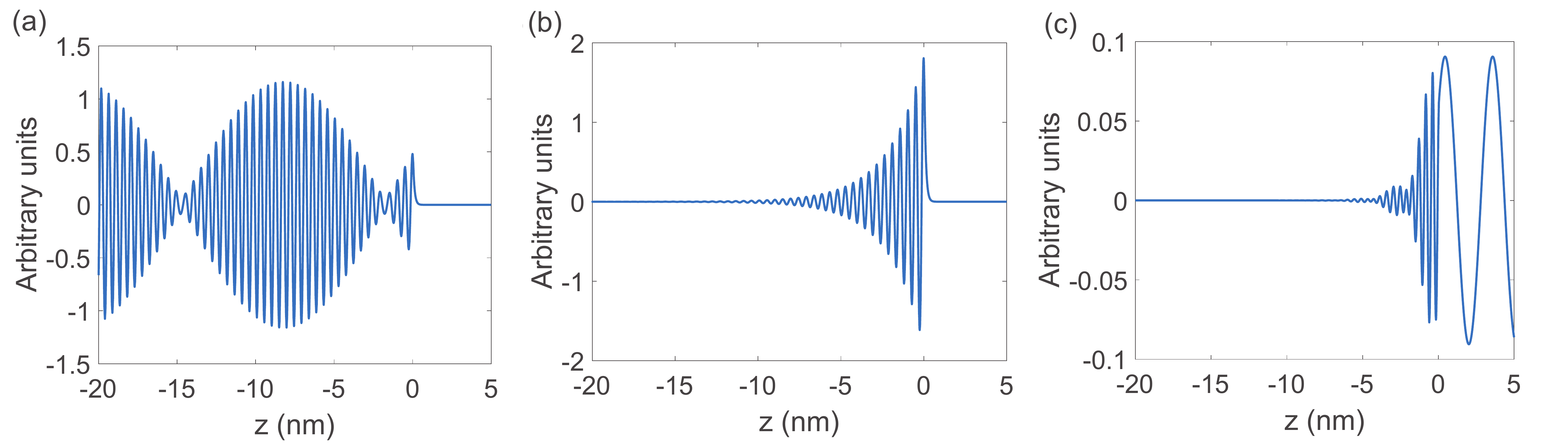}
\caption{Example of the real component of the $z$ dependent part of the initial bulk (a), surface (b) and final (c) wave functions. Here $z_0=0.042$ nm and $s=12.5$.
\label{fig2}}
\end{figure*}

\subsubsection{Initial Bulk States}

The $z$ dependent part of the initial wave functions for the bulk states inside the Ag(111) surface can be given by the combination of two Bloch states ($ k_{zi}+p$ and $ k_{zi}-p$) of the lower $sp$ band  and outside the surface can be given by an exponential decay\cite{Chaing2000}. Thus for $z<z_0$
\begin{equation}
\begin{split}
 \phi _{zi}=&N\bigg\{ e^{i\left( k_{zi}+p\right)z
}+\psi _{i}e^{i\left( k_{zi}-p\right)z } \\
& +c_{1}\left[e^{-i\left( k_{zi}+p\right)z}+\psi_{i}e^{-i\left(
k_{zi}-p\right)z }\right] \bigg\} 
\end{split}
\end{equation}
and for $z\ge z_0$
\begin{equation}
 \phi _{zi}=Nc_2e^{-\kappa_iz^{\prime}}
\end{equation}
 where $\kappa_i=\sqrt{2m_e\left(W-E_{zi}\right)}/\hbar$. 
 
The normalization constant $N$ can be obtained by normalizing the wave function. Constants $c_1$ and $c_2$ are obtained by matching the wave function and its derivative at $z=z_0$. The expressions for $\psi_i$, $N$, $c_1$ and $c_2$ are given in the appendix.

 \subsubsection{Initial Surface States} 
For the surface state $E_{zi}=E_s$, $k_{zi}=k_{zs}$, $\phi_{zi}=\phi_{zs}$ and $\psi _{i}=\psi _{s}$. The energy of the surface state ($E_s$) lies withing the $L$ gap. Hence, from equation \ref{dispersion} we see that the value of $k_{zs}$ is imaginary causing the surface state to decay into the bulk. 

For $z<z_0$ the wave function of the surface state (ss)  is given by
\begin{equation}
\begin{split}
 \phi _{zs}=&N_s\left[ e^{i\left( k_{zs}+p\right)z}+\psi _{s}e^{i\left( k_{zs}-p\right)z }\right]
\end{split}
\end{equation}
and for $z>z_0$ it is given by
\begin{equation}
 \phi _{zs}=N_sc_se^{-\kappa_sz^{\prime}}
\end{equation}
 where $\kappa_s=\sqrt{2m_e\left(W-E_{s}\right)}/\hbar$. 
$N_s$ is the normalization constant and can be obtained by normalizing the wave function. $c_s$ and $z_0$ can be obtained by matching the wave function and its derivative at $z=z_0$. Expressions for $N_s$ and $c_s$ are given in the appendix. An explicit expression cannot be obtained for $z_0$ and its value needs to be calculated numerically to satisfy the continuity conditions of the wave functions as given in the appendix.

\subsubsection{Final States}
The final state wave functions are not the free electron wave functions of the emitted electron, but are time reversed LEED states as required by the one step photoemission theory\cite{Mahan1970,Kevan1992}. Inside the Ag(111) surface they can be given by the combination of two Bloch states($ k_{zi}+p$ and $ k_{zi}-p$) of the upper $sp$ band along with an exponential decay to account for the various scattering mechanisms that prevent emission of excited electrons. The final wave functions outside the surface are plane waves. Thus for $z<z_0$

\begin{equation}
\phi _{zf}^{\ast }=t_{pk}^{\ast }\left\{ e^{\left[ -i\left(
k_{zf}+p\right) +k_{d}\right] z }+\psi _{f}e^{\left[
-i\left( k_{zf}-p\right) +k_{d}\right] z }\right\}\sqrt{\frac{2}{L}}
\end{equation}
where $k_d=s/d_{e-e}$ is the exponential decay constant that takes into account the scattering mechanisms that prevent emission of excited electrons. $d_{e-e}$ is the electron-electron scattering length, which is the dominant scattering mechanism in metals. The scattering parameter $s$ is used as a fitting parameter in the calculation.

For $z\ge z_0$
\begin{equation}
 \phi _{zf}^{\ast }=\left[e^{-ik_{z}z^{\prime} }+r_{pk}^{\ast
}e^{ik_{z}z^{\prime} }\right]\sqrt{\frac{2}{L}}
\end{equation}
where $\hbar k_z=\sqrt{2m_e\left(E_{fz}-W\right)}$ is the momentum of the emitted electron in the $z$ direction. Constants $t_{pk}^{\ast }$ and $r_{pk}^{\ast}$ are obtained by matching the wave function and its derivative at $z=z_0$. Expressions for $\psi_f$, $t_{pk}^{\ast }$ and $r_{pk}^{\ast}$ are given in the appendix. Note that the normalization of the final states is such that the out going plane wave representing the emitted photoelectron is normalized to unity.

Figure \ref{fig2} shows an example of the real component of the $z$ dependent part of the initial bulk, surface and final wave functions.

\subsection{Calculation of the Matirx Elements}
For $p$ polarized light, using the Hamiltonian from equation \ref{hamiltonian}, the matrix elements $M=\left\vert \left\langle \phi _{f}\left\vert \mathcal{H}\right\vert \phi _{i}\right\rangle \right\vert$ from equation \ref{trate1} can be written as
 \begin{equation}
\begin{split}
&M=\\ &\frac{e\hbar \left\vert A_0 T_p\right\vert}{m_ec}
\left\vert\left\langle\phi _{f1}\left\vert \sin \theta_{t}\frac{\partial }{\partial z}+\cos \theta_{t}\frac{\partial }{\partial x}+\sin \theta_{t}C\delta \left( z^{\prime}\right) \right\vert \phi_{i}\right\rangle \right\vert\label{mat_ele_p}
\end{split}
\end{equation} 
where $\phi_{f1}=\phi_f e^{\frac{z^{\prime }H\left(-z^{\prime }\right)}{d_l}}$  
      
Using the wave functions given in the previous section and integrating over the box one can calculate $M^2$ for $p$ polarized light as
\begin{equation}
\begin{split}
&M^2=\\ &\frac{4K_1\left\vert T_p\right\vert^2}{L^4}k_z^2 \left\vert\left(I_d+CI_s\right)\sin \theta_{t}+ik_{xi}I\cos \theta_{t}\right\vert^2\\&\times h\left(k_{xi}-k_{xf}\right)h\left(k_{yi}-k_{yf}\right)\label{m2_p}      
\end{split}
\end{equation}
where $K_1=\frac{e^2\hbar^2\left\vert A_0\right\vert^2}{m_e^2c^2}$ and $h\left(\zeta\right)=\frac{1}{L}\left[\frac{2\sin\left(L\zeta/2\right)}{\zeta}\right]^2$. Note that in the limit $L\rightarrow\infty$, $h\left(\zeta\right)=2\pi\delta\left(\zeta\right)$. $I_d$, $I_s$ and $I$ are given as follows         
\begin{eqnarray}
I_d&=&\frac{L/2}{k_z}\int_{-L/2}^{L/2} dz\phi _{zf}^{\ast }e^{\frac{z^{\prime }H\left(-z^{\prime }\right)}{d_l}}\frac{\partial }{\partial z}\phi _{zi}\\
I_s&=&\frac{L/2}{k_z}\phi_{zf}^{\ast }\left( z_{0}\right) \phi _{zi}\left( z_{0}\right)\\
I&=&\frac{L/2}{k_z}\int_{-L/2}^{L/2} dz\phi _{zf}^{\ast }e^{\frac{z^{\prime }H\left(-z^{\prime }\right)}{d_l}}\phi_{zi}
\end{eqnarray}
The integrals $I_d$ and $I$ can be evaluated analytically and the expressions are given in the appendix. Owing to the appropriate normalization of the wave functions, $I_d$,$I_s$ and $I$ are independent of $L$ when $L\rightarrow\infty$. 

Note that the matrix element given in equation \ref{m2_p} is asymmetric in $k_{xi}$. This can lead to an asymmetric photoemission where the number of electrons emitted with momentum $k_x$ is different from electrons with number of electrons with $x$ direction momentum $-k_x$. 

The matrix element for $s$ polarized light can be given by
\begin{equation}
\begin{split}
&M^2=\\ &\frac{4K_1\left\vert T_s\right\vert^2}{L^4}k_z^2 \left\vert i k_{yi}I\right\vert^2 h\left(k_{xi}-k_{xf}\right)h\left(k_{yi}-k_{yf}\right)      
\end{split}
\end{equation}

\subsection{Calculating the QE}

The total QE can be written as sum of QE contribution from the bulk states ($\mathrm{QE}_{\mathrm{bulk}}$) and the surface state ($\mathrm{QE}_{\mathrm{ss}}$)
\begin{equation}
\mathrm{QE}=\mathrm{QE}_{\mathrm{bulk}}+\mathrm{QE}_{\mathrm{ss}}.
\end{equation}

$\mathrm{QE}_{\mathrm{bulk}}$ can be given by
\begin{equation}
\mathrm{QE}_{\mathrm{bulk}}=\frac{R_\mathrm{bulk}}{FL^2}\label{QE_bulk}
\end{equation}
where $R_\mathrm{bulk}$ can be calculated from equation \ref{trate1} with the integrations being carried over all possible initial bulk and final states. $F$ is the incident photon flux per unit area given by equation \ref{flux}.

$\mathrm{QE}_{\mathrm{ss}}$ can be given by
\begin{equation}
\mathrm{QE}_{\mathrm{ss}}=\frac{R_\mathrm{ss}}{FL^2}.\label{QE_ss}
\end{equation}

$R_\mathrm{ss}$ can be calculated by an expression similar to equation \ref{trate1} with the difference that the integration over the initial state have to be performed over $k_{xi}$ and $k_{yi}$ only, due to the 2-D nature of the surface state. $R_\mathrm{ss}$ can be written as 
\begin{equation}
\begin{split}
&R_\mathrm{ss}=\\ &\frac{4\pi}{\hbar}\left(\frac{L}{2\pi}\right)^5\int\int dk_{xi} dk_{yi}\int d^3\vec{k}M ^{2}\delta\left(E_f-\left(E_i+\hbar\omega\right)\right)F\left(E_i\right)\label{trate1ss}
\end{split}
\end{equation}
  
Using equations \ref{flux}, \ref{trate1},  \ref{delta}, \ref{m2_p} and \ref{QE_bulk},  $\mathrm{QE}_{\mathrm{bulk}}$ for $p$ polarized light can be written as
\begin{widetext}
\begin{equation}
\mathrm{QE}_{\mathrm{bulk}}=\frac{K}{\left(2\pi\right)^2}\int d^3\vec{k_i}\int d^3\vec{k} \frac{k_z^2}{k_{z0}} \left\vert\left(I_d+CI_s\right)\sin \theta_{t}+ik_{xi}I\cos \theta_{t}\right\vert^2\times h\left(k_{xi}-k_{x}\right)h\left(k_{yi}-k_{y}\right)\delta\left(k_z-k_{z0}\right)F\left(E_i\right)\label{QE_expr1}
\end{equation}
\end{widetext}
where $K=\frac{8\left(\hbar c\right)^2\alpha\left\vert T_p\right\vert^2}{\left(2\pi\right)^2\left(m_ec^2\right)\hbar\omega}$ and $\alpha$ is the fine structure constant. Note that we require $k_{xf}=k_x$ and $k_{yf}=k_y$ in order to match the transverse part of the final wave functions at the boundary.

In the limit as $L\rightarrow\infty$, $h\left(\zeta\right)=2\pi\delta\left(\zeta\right)$. Taking the limit as $L\rightarrow\infty$ and integrating over the final states we obtain
\begin{equation}
\begin{split}
&\mathrm{QE}_{\mathrm{bulk}}=\\&K\int d^3\vec{k_i}k_{z0}\left\vert\left(I_d+CI_s\right)\sin \theta_{t}+ik_{xi}I\cos \theta_{t}\right\vert^2\times F\left(E_i\right)
\end{split}
\end{equation}
Note that $I$, $I_d$, $I_s$, $k_z0$ and $E_i$ are functions of $\vec{k_i}$ and $\vec{k}$. Integrating the $\delta$ functions in equation \ref{QE_expr1} we get $k_x=k_{xi}$, $k_y=k_{yi}$ and $k_z=k_{z0}$. These $\delta$ functions enforce the conservation of transverse momentum and energy during photoemission.

Similarly, $\mathrm{QE}_{\mathrm{ss}}$ can be obtained using \ref{QE_ss} as
\begin{equation}
\begin{split}
&\mathrm{QE}_{\mathrm{ss}}=\frac{2\pi K}{L}\times\\&\int d{k_{xi}}\int d{k_{yi}}k_{z0}\left\vert\left(I_d+CI_s\right)\sin \theta_{t}+ik_{xi}I\cos \theta_{t}\right\vert^2\times F\left(E_i\right)\label{QE_expr2}
\end{split}
\end{equation}
The normalization constant for the surface state ($N_s$) is not dependent on $L$. Hence, for the surface state $I$, $I_d$ and $I_s$ are proportional to $sqrt(L)$ even as $L\rightarrow\infty$. Thus, the surface state QE as given in equation \ref{QE_expr2} remains independent of $L$ as $L\rightarrow\infty$.

After writing $k_{xi}$ and $k_{yi}$ in cylindrical co-ordinates as $k_{xi}=k_r\cos\varphi$ and $k_{xi}=k_r\sin\varphi$; then integrating over $\varphi$ the above expressions for $\mathrm{QE}_{\mathrm{bulk}}$ and $\mathrm{QE}_{\mathrm{ss}}$ can be written as
\begin{equation}
\begin{split}
&\mathrm{QE}_{\mathrm{bulk}}=2\pi K\int dk_r\int dk_{zi}k_rk_{z0}F\left(E_i\right)\times\\&\left[\left\vert\left(I_d+CI_s\right)\sin\theta_{t}\right\vert^2+k_{r}^2\frac{\left\vert I\cos\theta_{t}\right\vert^2}{2}\right]\label{QE_expr3}
\end{split}
\end{equation}
 and
\begin{equation}
\begin{split}
&\mathrm{QE}_{\mathrm{ss}}=\frac{4\pi^2K}{L}\int dk_rk_rk_{z0}F\left(E_i\right)\times\\&\left[\left\vert\left(I_d+CI_s\right)\sin\theta_{t}\right\vert^2+k_{r}^2\frac{\left\vert I\cos\theta_{t}\right\vert^2}{2}\right]\label{QE_expr4} 
\end{split}
\end{equation}
respectively.

The QE for $s$ polarized light can be similarly calculated by using the appropriate matrix elements. The 3-D momentum distributions and the rms transverse momentum can also be calculated easily as shown in equation \ref{rmsmom}.

\section{Results and Discussion}
\subsection{Spectral response}
\begin{figure}
\includegraphics[width=3.2in]{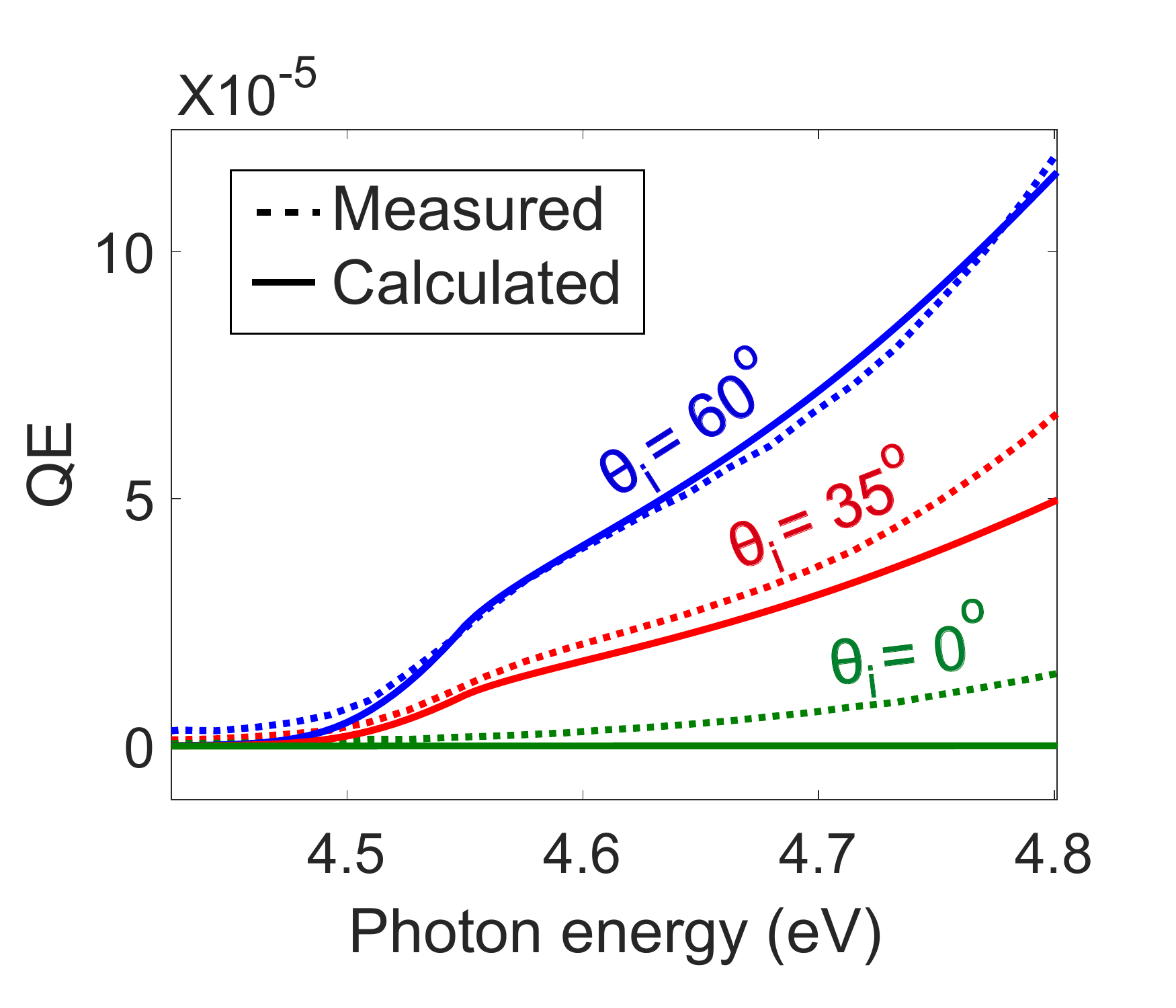}
\caption{Measured and calculated spectral response of the Ag(111) surface at various angles of incidence ($\theta_i$) in $p$ polarized light. The error bar on the experimental measurement is $\sim$ 10\%
\label{fig3}}
\end{figure}

Figure \ref{fig3} compares the spectral response measured from an Ag(111) surface to the result obtained from the photoemission model presented above, for $p$ polarized light at various angles of incidence.

In order to measure the QE, a commercially bought\cite{spl} single crystal Ag(111) sample was prepared in an ultra-high vacuum chamber with base pressure in the low $10^{-10}$ torr range. Several cycles of Ar ion bombardment and annealing to $500^{\circ}$C were performed until a sharp hexagonal LEED pattern was observed. The surface cleanliness was verified using Auger electron spectroscopy. The QE was obtained by measuring the photocurrent and the power of light incident on the sample surface. A laser based plasma lamp with a monochromator\cite{Feng2013} was used as a light source for the QE measurement. The spectral width of the light source was 2 nm FWHM. 

All constants used for modeling the band structure to calculate the QE are given in table \ref{tab1}. The optical constants ($n_r$ and $n_i$) for silver as a function of wavelength are well known\cite{optical_const1,optical_const2}. The surface constant $C$ and the electron-electron scattering length $d_{e-e}$ were obtained as a function of photon energy by extrapolating the values of $C$ and $d_{e-e}$ obtained from PES measurements\cite{Miller1996,Miller1997}. The scattering parameter $s$ is set to 12.5 to obtain a good match to the experimental data. 

Figure \ref{fig3} shows that the calculated QE explains the experimental data, both qualitatively and quantitatively. With the exception of the scattering parameter $s$, this photoemission model calculates the QE accurately without the use of any ad hoc coefficients or scaling factors.  
It is seen that the QE increases with the angle of incidence for $p$ polarized light (vectorial photoelectric effect). The knee observed in the spectral response for higher angles of incidence at $\sim$ 4.55 eV is caused due to the surface state. This becomes clear from figure \ref{fig4} which shows the contributions to the QE from the bulk and surface states.   
The sections below discuss the effect of the scattering parameter and the vectorial photoelectric effect respectively.

\subsubsection{Effect of Scattering}
\begin{figure}
\includegraphics[width=3.2in]{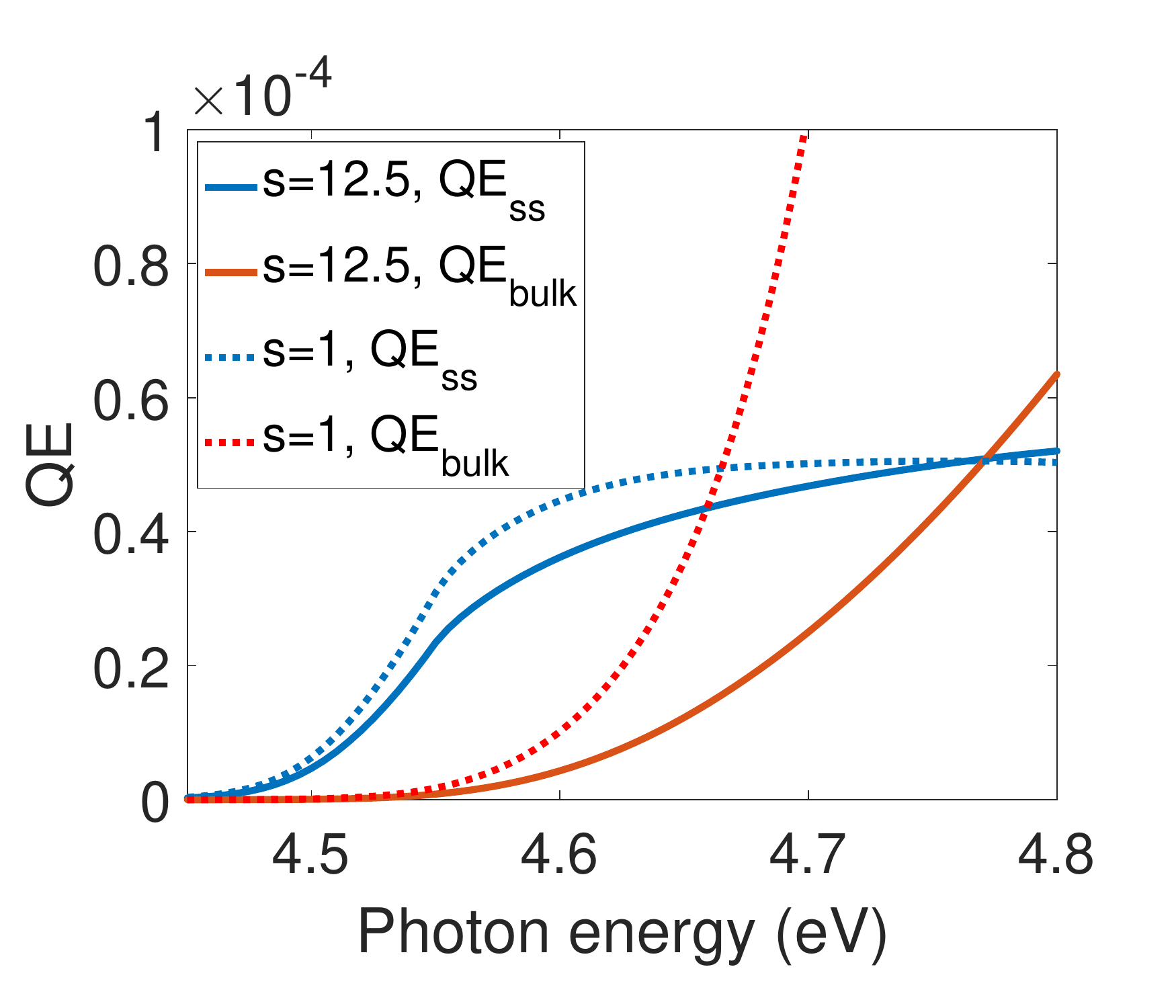}
\caption{$\mathrm{QE}_{\mathrm{bulk}}$ and $\mathrm{QE}_{\mathrm{ss}}$ for $s=1$ and $s=12.5$. $\mathrm{QE}_{\mathrm{ss}}$ is not very sensitive to $s$ because the surface state electrons are localized at the surface and do not need to travel inside the metal to get emitted.
\label{fig4}}
\end{figure}
The decay constant of the final wave function $k_d$ takes into account the electrons that were excited by light but were unable to escape due to various scattering mechanisms while travelling towards the surface. The inelastic electron-electron scattering is the dominant scattering mechanism of excited electrons in metals. Hence we write $k_d=s/d_{e-e}$, where $d_{e-e}$ is the electron-electron scattering mean free path and $s$ is adjusted to match the calculated QE to the experimental value.
Figure \ref{fig4} shows $\mathrm{QE}_{\mathrm{bulk}}$ and $\mathrm{QE}_{\mathrm{ss}}$ for $s=1$ and $s=12.5$. We can see that $\mathrm{QE}_{\mathrm{ss}}$ does not change significantly with $s$. The surface state is localized at the metal-vacuum interface. Hence the electrons excited from the surface state do not need to travel inside the metal to get emitted. This causes $\mathrm{QE}_{\mathrm{ss}}$ to be insensitive to $s$ or $k_d$. 

In order to match the experimental data, $s$ needs to be set to a particularly large value of 12.5. This implies a much higher effective scattering rate than set by the electron-electron scattering lengths obtained from UV-PES data\cite{Miller1996,Miller1997}. The reason for this increased scattering is not clear.

\subsubsection{Vectorial Photoelectric Effect}
\begin{figure}
\includegraphics[width=3.2in]{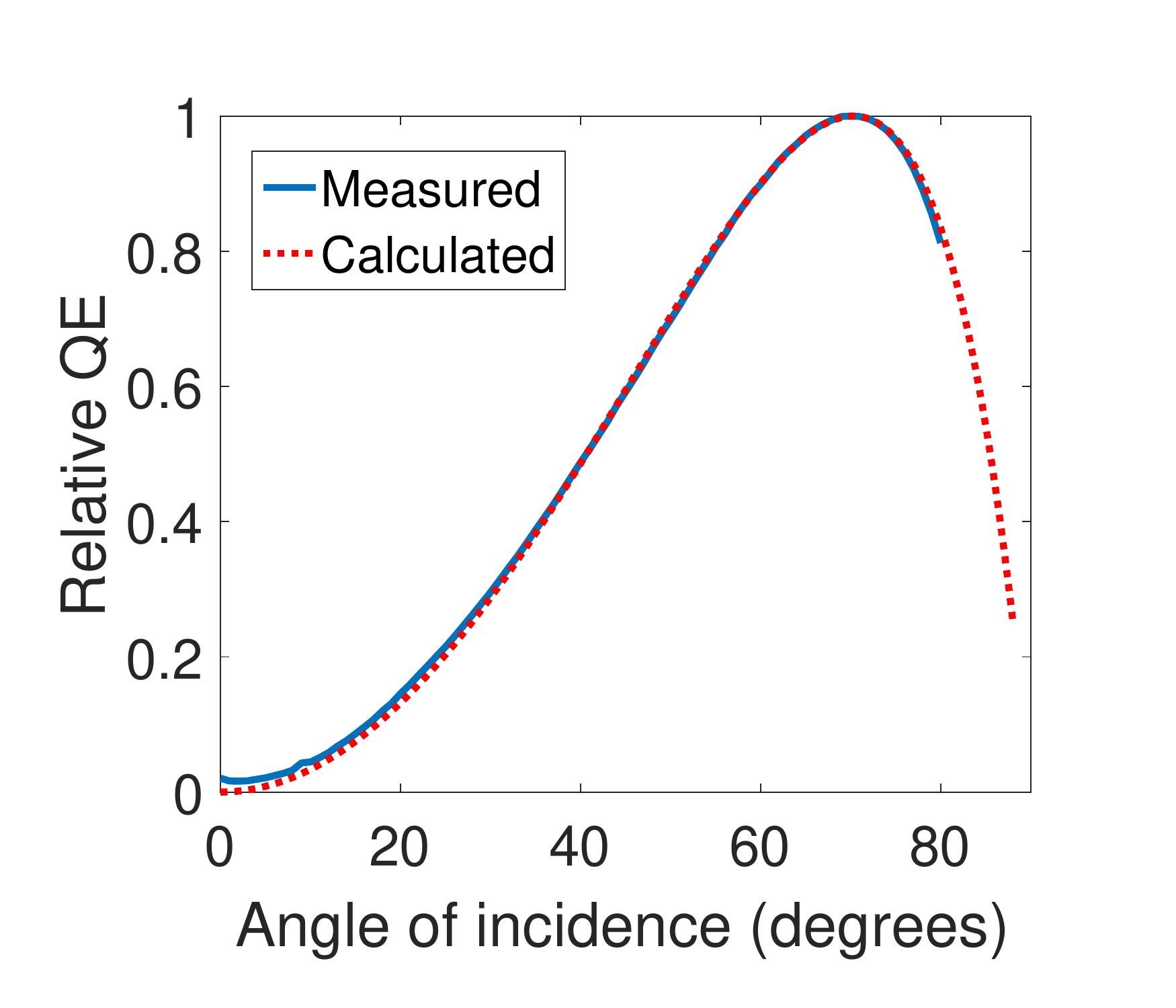}
\caption{The theoretical curve for the vectorial photoelectric effect obtained from equation \ref{VPE}  matches the experimental data measured at photon energy of 4.57 eV
\label{fig5}}
\end{figure}

Vectorial photoelecric effect is the variation of QE with the angle of incidence and polarization of incident light.

The QE for $p$ polarized light is given by equations \ref{QE_expr3} and \ref{QE_expr4}. In these equations, the term $\left\vert\left(I_d+CI_s\right)\sin\theta_{t}\right\vert^2$ corresponds to the QE contribution of the $z$ component of the polarization vector and the term $k_{r}^2\frac{\left\vert I\cos\theta_{t}\right\vert^2}{2}$ corresponds to the QE contribution of the $x$ component of the polarization vector. For the band structure and wave functions used here, $\left\vert\left(I_d+CI_s\right)\right\vert^2\gg k_{r}^2\frac{\left\vert I\right\vert^2}{2}$. As a result, the photoemission from the Ag(111) surface is dominated by the $z$ component of the polarization vector (i.e component perpendicular to the surface). Neglecting the contribution of the $x$ (parallel to surface) component, the QE can be written as 
\begin{equation}
\mathrm{QE}=K_p\frac{\left\vert T_p\sin\theta_{t}\right\vert^2}{\cos\theta_i} \label{VPE}
\end{equation}
where $K_p$ is a constant independent of the angle of incidence. Note that both $T_p$ and $\theta_{t}$ are dependent on the angle of incidence. Figure \ref{fig5} shows that the experimentally measured angular dependence of QE for $p$ polarized light matches this calculation. This dependence is similar to the angular dependence of QE measured for several materials\cite{Broudy1971,Pedersoli2008,Benemanskaya1994,Juenker1964}.

The spectral response calculated by the model at $0^{\circ}$ angle of incidence is much smaller than the experimental value (see figure \ref{fig3}). At $0^{\circ}$ angle of incidence only the $x$ and $y$ components of the polarization vector exist.  This implies that the experimentally observed contribution of the $x$ and $y$ components of of the polarization vector is larger than that calculated by the model. The assumption that the wave functions in the $x$ and $y$ directions are modeled by plane waves could be one possible culprit for this. Emission from parts of the band structure not modeled by the nearly free electron representation, many body photoemission effects like the hole state lifetime induces energy spread\cite{hufner_width} and the breakdown of the sudden approximation\cite{sudden_approx} are other effects which may be responsible for this discrepancy. They may also be responsible for the large effective scattering parameter.


\subsection{Transverse Momentum Spread}
\begin{figure}
\includegraphics[width=3.2in]{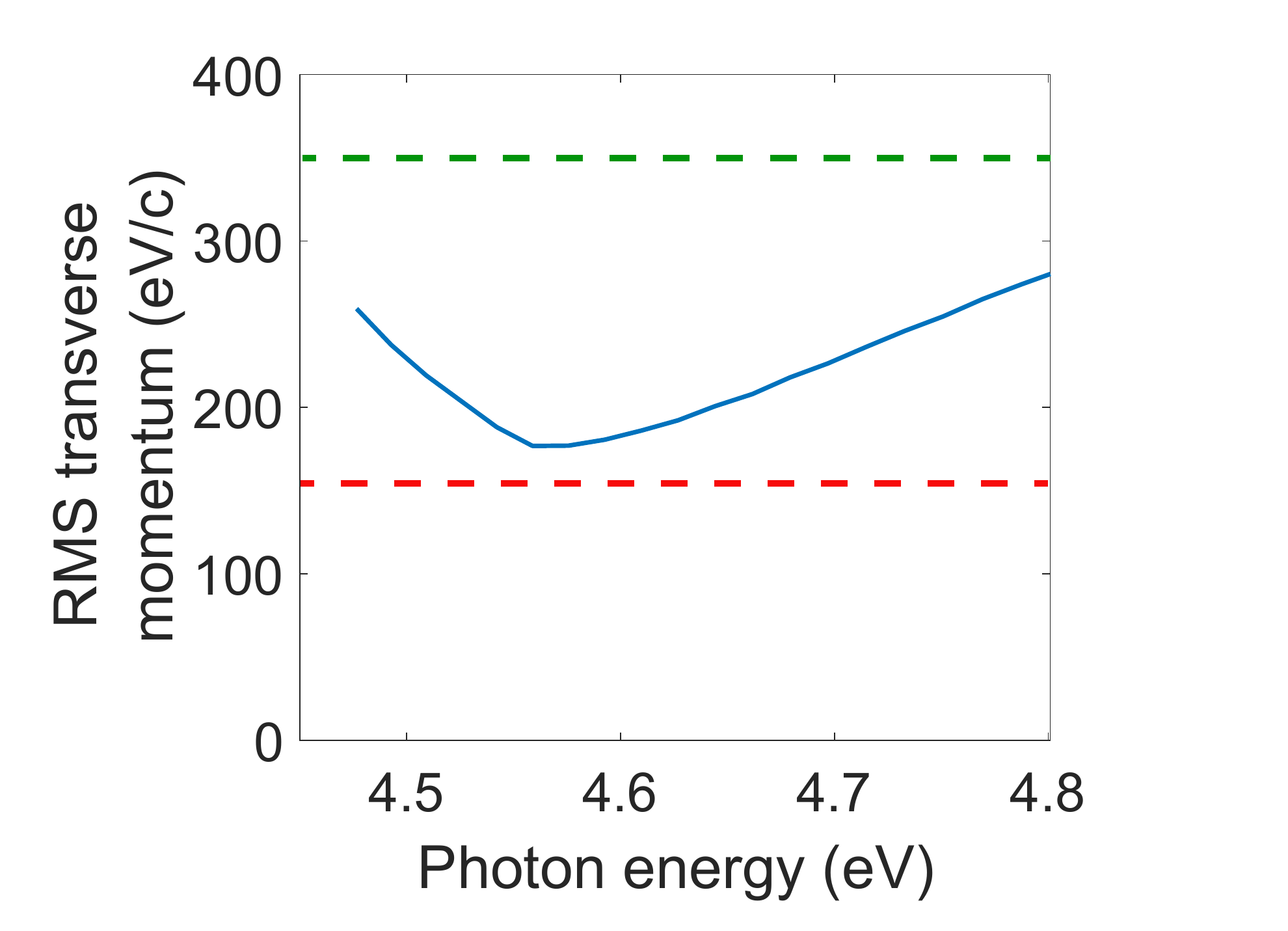}
\caption{The blue curve shows the rms transverse momentum calculated for the Ag(111) surface. The dashed green line shows the rms transverse momentum typically obtained from cathodes in the state-of-art electrons guns. The red dashed line indicated the thermally limited rms transverse momentum\cite{Jun_thermal}. At 4.57 eV photon energy Ag(111) gives near thermal rms transverse momentum with a QE of $5\times10^{-5}$ making it a much better electron source than the current state-of-art. 
\label{fig6}}
\end{figure}
Figure \ref{fig6} shows the rms transverse momentum expected from the Ag(111) surface. The rms transverse momentum has been calculated using equation \ref{rmsmom} for angle of incidence equal to 60$^{\circ}$ . It can be seen that the rms transvserse momentum initially decreases , reaches a minimum and then increases with increasing photon energy. At photon energies very close to threshold only electrons from the ring formed by the intersection of the surface state with the Fermi level are emitted. These electrons have a relatively high transverse momentum. At higher photon energies, electrons from the surface state with lower transverse momentum and lower energy can also be emitted along with the electrons from the surface state ring at the Fermi level. This causes the rms transverse momentum to initially reduce with increasing photon energy. This decline continues till the photon energy is sufficiently high to allow emission from the entire surface state. At this photon energy the rms momentum reaches a minimum. At higher photon energies, the electrons from bulk states which are located near the Fermi level and have a much higher transverse momentum are allowed to be emitted causing the rms transverse momentum to increase again.

The smallest rms momentum measured from polycrystalline metal cathodes (which are typically used as electron sources) is limited by the room temperature to a value of $160$ eV/c\cite{Jun_thermal} at the photoemission threshold. However, at the photoemission threshold the QE is also very low (in the $10^{-6}$-$10^{-7}$) range making polycrystalline cathodes unusable in this regime. At higher photon energies, the QE increases but so does the rms transverse momentum. In order to obtain a desirable QE greater than 10$^{-5}$ the photon energy used has to be several 100 meV above threshold. This sets the rms transverse momentum practically obtained in electron guns to $\sim350$ eV/c\cite{Dowell}. According to our calculations, Ag(111) when operated at an angle of incidence of 60$^{\circ}$ in $p$ polarized light at 4.57 eV can act as a cathode with rms transverse momentum lower than 180 eV/c and a QE as high as $5\times10^{-5}$. This shows that Ag(111) can act as a better photocathode the polycrystalline metals currently used as electron sources.    

\section{Conclusion}
We have presented a 3-D one step model that allows us to calculate photoemission properties like QE and rms transverse momentum of emitted electrons from single crystal surfaces. Optimizing these photoemission properties can greatly improve the performance of electron source applications like FELs and UED.

Using the example of photoemission from Ag(111) we show that not only can this model calculate the spectral response from surface state without the use of any ad hoc parameters, but also explains the photoemission phenomena of the vectorial photoelectric effect accurately.

We also calculate the rms transverse momentum from an Ag(111) surface and show that in $p$ polarized light with a high angle of incidence, the Ag(111) surface can exhibit high QE along with a small rms transverse momentum, making it a much better cathode than the currently used polycrystalline metals. Upon integrating with other band structure and wave function calculation techniques like density functional theory, this methodology can be used to calculate the electron source relevant photoemission properties from any single crystal surface in order to identify ideal electron emitters from first principle calculations\cite{harrison}. Such a methodology is essential to screen for materials to identify good electron emitters.

\section{Acknowledgements}
The authors S. Karkare and W. Wan have contributed equally to this work.
The authors would like to thank Dr. T. Miller for stimulating
discussions. This work was supported by the Director, Office of Science,
Office of Basic Energy Sciences of the U. S. Department of Energy, under
Contract Nos. KC0407-ALSJNT-I0013 and DE-AC02-05CH11231 (W. W., S. K., J.
F., H. A. P.) and the U.S. National Science Foundation under Grant No. DMR
13-05583 (T.-C. C.).

\begin{widetext}
\section{Appendix}

The analytic expressions to calculate several of the coefficients used in the wave function calculations are given below
\begin{eqnarray}
\psi_{i,f}&=&\frac{\left(\frac{\hbar^2\left(k_{zi,zf}+p\right)^2}{2m_{i,f}}-\frac{\hbar^2p^2}{2m_{i,f}}-E_{zi,zf}+V+E_v\right)}{V}\\
c_1&=&\frac{\left[\kappa_i+i\left(k_{zi}+p\right)\right]e^{i\left(k_{zi}+p\right)z_0}+\left[\kappa_i+i\left(k_{zi}-p\right)\right]\psi_ie^{i\left(k_{zi}-p\right)z_0}}{\left[-\kappa_i+i\left(k_{zi}+p\right)\right]e^{-i\left(k_{zi}+p\right)z_0}+\left[-\kappa_i+i\left(k_{zi}-p\right)\right]\psi_ie^{-i\left(k_{zi}-p\right)z_0}}\\
c_2&=&2i\frac{\left(k_{zi}+p\right)+\left(k_{zi}-p\right)\psi_i^2+2k_{zi}\psi_i\cos\left(2pz_0\right)}{\left[\kappa_i+i\left(k_{zi}+p\right)\right]e^{-i\left(k_{zi}+p\right)z_0}+\left[-\kappa_i+i\left(k_{zi}-p\right)\right]e^{-i\left(k_{zi}-p\right)z_0}}\\
t_{pk}&=&\frac{2ik_z}{\left[i\left(k_{zf}+p+k_z\right)+k_d\right]e^{\left[i\left(k_{zf}+p\right)+k_d\right]z_0}+\left[i\left(k_{zf}-p+k_z\right)+k_d\right]\psi_fe^{\left[i\left(k_{zf}-p\right)+k_d\right]z_0}}\\
1+r_{pk}&=&\frac{2ik_z\left[e^{\left[i\left(k_{zf}+p\right)+k_d\right]z_0}+\psi_fe^{\left[i\left(k_{zf}-p\right)+k_d\right]z_0}\right]}{\left[i\left(k_{zf}+p+k_z\right)+k_d\right]e^{\left[i\left(k_{zf}+p\right)+k_d\right]z_0}+\left[i\left(k_{zf}-p+k_z\right)+k_d\right]\psi_fe^{\left[i\left(k_{zf}-p\right)+k_d\right]z_0}}\\
N&=&\frac{\sqrt(2)}{\sqrt{\left(\frac{L}{2}\right)\left(1+\psi_i^2\right)\left(1+\left\vert c_1\right\vert^2\right)}}\left(\mathrm{assuming} L\rightarrow\infty\right)\\
c_s&=& e^{i\left(k_{zs}+p\right)z_0}+\psi_se^{i\left(k_{zs}-p\right)z_0}\\
N_s&=&\sqrt(2)\left[\frac{e^{2k_{zs}z_0}}{2k_{zs}}\left(1+\left\vert \psi_s\right\vert^2\right)+\mathrm{Re}\left[\psi_s^{\ast}\frac{e^{2\left(ip+k_{zs}\right)}}{ip+k_{zs}}\right]+\frac{\left\vert c_s\right\vert^2}{2\kappa_s}\right]^{-\frac{1}{2}} \\
\end{eqnarray}
Note that the wave functions have been normalized to 2 in order to account for the electrons emitted from the equivalent $L$ point at $\left(-\frac{\pi}{a},-\frac{\pi}{a},-\frac{\pi}{a}\right)$.
$z_0$ can be obtained by solving the following equation numerically
\begin{equation}
i\left(k_{zs}+p\right)e^{i\left(k_{zs}+p\right)z_0}+i\psi_s\left(k_{zs}-p\right)e^{i\left(k_{zs}-p\right)z_0}=-\kappa_sc_s
\end{equation}

The analytic expressions for the integrals $I_d$ and $I$ used in the matrix element calculations are given below

\begin{eqnarray}
I_d=\int dz\phi _{zf}^{\ast }\frac{\partial }{\partial z}\phi _{zi} &=&N\kappa
_{i}c_{2}\left( \frac{2ik_{z}}{\kappa _{i}^{2}+k_{z}^{2}}+\frac{%
1+r_{pk}^{\ast }}{-\kappa _{i}+ik_{z}}\right)  \\
&&+Nt_{pk}^{\ast }\left[ \frac{i\left( k_{zi}+p\right) e^{\left[ i\left(
-k_{zf}+k_{zi}\right) +k_{d}\right] z_{0}}}{i\left( -k_{zf}+k_{zi}\right)
+k_{d}}+\phi _{i}\frac{i\left( k_{zi}-p\right) e^{\left[ i\left(
-k_{zf}+k_{zi}-2p\right) +k_{d}\right] z_{0}}}{i\left(
-k_{zf}+k_{zi}-2p\right) +k_{d}}\right.  \\
&&-c_{1}\frac{i\left( k_{zi}+p\right) e^{\left[ -i\left(
k_{zf}+k_{zi}+2p\right) +k_{d}\right] z_{0}}}{-i\left(
k_{zf}+k_{zi}+2p\right) +k_{d}}-c_{1}\phi _{i}\frac{i\left( k_{zi}-p\right)
e^{\left[ -i\left( k_{zf}+k_{zi}\right) +k_{d}\right] z_{0}}}{-i\left(
k_{zf}+k_{zi}\right) +k_{d}} \\
&&+\phi _{f}\frac{i\left( k_{zi}+p\right) e^{\left[ i\left(
-k_{zf}+k_{zi}+2p\right) +k_{d}\right] z_{0}}}{i\left(
-k_{zf}+k_{zi}+2p\right) +k_{d}}+\phi _{i}\phi _{f}\frac{i\left(
k_{zi}-p\right) e^{\left[ i\left( -k_{zf}+k_{zi}\right) +k_{d}\right] z_{0}}%
}{i\left( -k_{zf}+k_{zi}\right) +k_{d}} \\
&&\left. -c_{1}\phi _{f}\frac{i\left( k_{zi}+p\right) e^{\left[ -i\left(
k_{zf}+k_{zi}\right) +k_{d}\right] z_{0}}}{-i\left( k_{zf}+k_{zi}\right)
+k_{d}}-c_{1}\phi _{i}\frac{i\left( k_{zi}-p\right) e^{\left[ -i\left(
k_{zf}+k_{zi}-2p\right) +k_{d}\right] z_{0}}}{-i\left(
k_{zf}+k_{zi}-2p\right) +k_{d}}\right] \\
I=\int dz\phi _{zf}^{\ast }\phi _{zi} &=&Nc_{2}\left( \frac{2ik_{z}}{\kappa
_{i}^{2}+k_{z}^{2}}+\frac{1+r_{pk}^{\ast }}{-\kappa _{i}+ik_{z}}\right)  \\
&&+Nt_{pk}^{\ast }\left[ \frac{e^{\left[ i\left( -k_{zf}+k_{zi}\right) +k_{d}%
\right] z_{0}}}{i\left( -k_{zf}+k_{zi}\right) +k_{d}}+\phi _{i}\frac{e^{%
\left[ i\left( -k_{zf}+k_{zi}-2p\right) +k_{d}\right] z_{0}}}{i\left(
-k_{zf}+k_{zi}-2p\right) +k_{d}}\right.  \\
&&-c_{1}\frac{e^{\left[ -i\left( k_{zf}+k_{zi}+2p\right) +k_{d}\right] z_{0}}%
}{-i\left( k_{zf}+k_{zi}+2p\right) +k_{d}}-c_{1}\phi _{i}\frac{e^{\left[
-i\left( k_{zf}+k_{zi}\right) +k_{d}\right] z_{0}}}{-i\left(
k_{zf}+k_{zi}\right) +k_{d}} \\
&&+\phi _{f}\frac{e^{\left[ i\left( -k_{zf}+k_{zi}+2p\right) +k_{d}\right]
z_{0}}}{i\left( -k_{zf}+k_{zi}+2p\right) +k_{d}}+\phi _{i}\phi _{f}\frac{e^{%
\left[ i\left( -k_{zf}+k_{zi}\right) +k_{d}\right] z_{0}}}{i\left(
-k_{zf}+k_{zi}\right) +k_{d}} \\
&&\left. -c_{1}\phi _{f}\frac{e^{\left[ -i\left( k_{zf}+k_{zi}\right) +k_{d}%
\right] z_{0}}}{-i\left( k_{zf}+k_{zi}\right) +k_{d}}-c_{1}\phi _{i}\frac{e^{%
\left[ -i\left( k_{zf}+k_{zi}-2p\right) +k_{d}\right] z_{0}}}{-i\left(
k_{zf}+k_{zi}-2p\right) +k_{d}}\right] ,
\end{eqnarray}%
\end{widetext}


%

\end{document}